\documentclass[reprint,superscriptaddress,showpacs,preprintnumbers,amsmath,amssymb,aps,pra]{revtex4-1}

\usepackage{mathtools}
\usepackage{graphicx}
\usepackage{subfigure}
\usepackage{multirow}
\renewcommand{\thesubfigure}{\alph{subfigure}}
\makeatletter
\renewcommand{\@thesubfigure}{(\thesubfigure)\space}
\makeatother
\usepackage{dcolumn}
\usepackage{bm}
\usepackage[colorlinks, linkcolor=blue, citecolor=blue, urlcolor=blue]{hyperref}
\usepackage{geometry}
\begin{document}

\title{Quantum phase transitions in a spin-1 antiferromagnetic chain \\
       with long-range interactions and modulated single-ion anisotropy}

\author{Jie Ren}
\affiliation{\mbox{School of Electronic and Information Engineering,
             Changshu Institute of Technology, Changshu 215500, China}}

\author{Wen-Long You}
\email{wlyou@nuaa.edu.cn}
\affiliation{College of Science, Nanjing University of Aeronautics and Astronautics,
             Nanjing 211106, China}
\affiliation{School of Physical Science and Technology, Soochow University, Suzhou,
             Jiangsu 215006, China}

\author {Andrzej M. Ole\'s$\,$}
\affiliation{\mbox{Institute of Theoretical Physics, Jagiellonian University,
             Profesora Stanis\l{}awa \L{}ojasiewicza 11, PL-30348 Krak\'ow, Poland}}
\affiliation{\mbox{Max Planck Institute for Solid State Research,
             Heisenbergstrasse 1, D-70569 Stuttgart, Germany} }

\date{May 19, 2020}

\begin{abstract}
We study the phase diagram of spin-1 antiferromagnetic chain with
isotropic antiferromagnetic interactions decaying with a power-law
$\propto r^{-\alpha}$ ($\alpha\ge 1$) accompanied by modulated
single-ion anisotropy. Employing the techniques of the density-matrix
renormalization group, effects of long-range interactions and
single-ion anisotropy on a variety of correlations are investigated.
In order to check the consistency, the fidelity susceptibilities are
evaluated across quantum phase transitions. The quantum critical
points are faithfully detected and orders of phase transitions are
determined. The correlation-length critical exponent is extracted from
scaling functions of the fidelity susceptibility. The presence of
long-range interactions leads to quantitative change of the phase
boundaries and reduces the order of phase transition under certain
conditions. A direct first-order transition between the periodic
N\'eel phase and the large-$D$ phase occurs for slowly decaying
antiferromagnetic interactions.
\end{abstract}

\maketitle

\section{introduction}
\label{sec:intorduction}

The Haldane phase~\cite{Haldane,Haldane1} of the one-dimensional (1D)
antiferromagnetic (AF) spin-1 ($S=1$) chain has received continued
attention as it is closely related to the breaking of a hidden
$\mathbb{Z}_2\otimes\mathbb{Z}_2$ symmetry \cite{Kennedy1992} through
a nonlocal unitary transformation. In fact, long-range order is absent
in the isotropic 1D spin-1 chain with short-range exchange interaction
but the ground state is characterized by a finite spectral gap and
exponentially decaying AF spin correlations. The thoroughly studied
gapped phase was conjectured by Haldane~\cite{Haldane,Haldane1},
and was confirmed in a series of experimental and theoretical papers
\cite{Nijs,Boschi,Nomura,Buyers,Malvezzi2016}. An important benchmark
of the Haldane phase is the occurrence of the nonlocal string order
in the isotropic AF $S=1$ Heisenberg chain \cite{Whi93}. At the same
time, the string order is a manifestation of the topological hidden
order, as pointed out by Kennedy and Tasaki \cite{Kennedy1992}.

\begin{figure}[t]
\includegraphics[width=1\columnwidth]{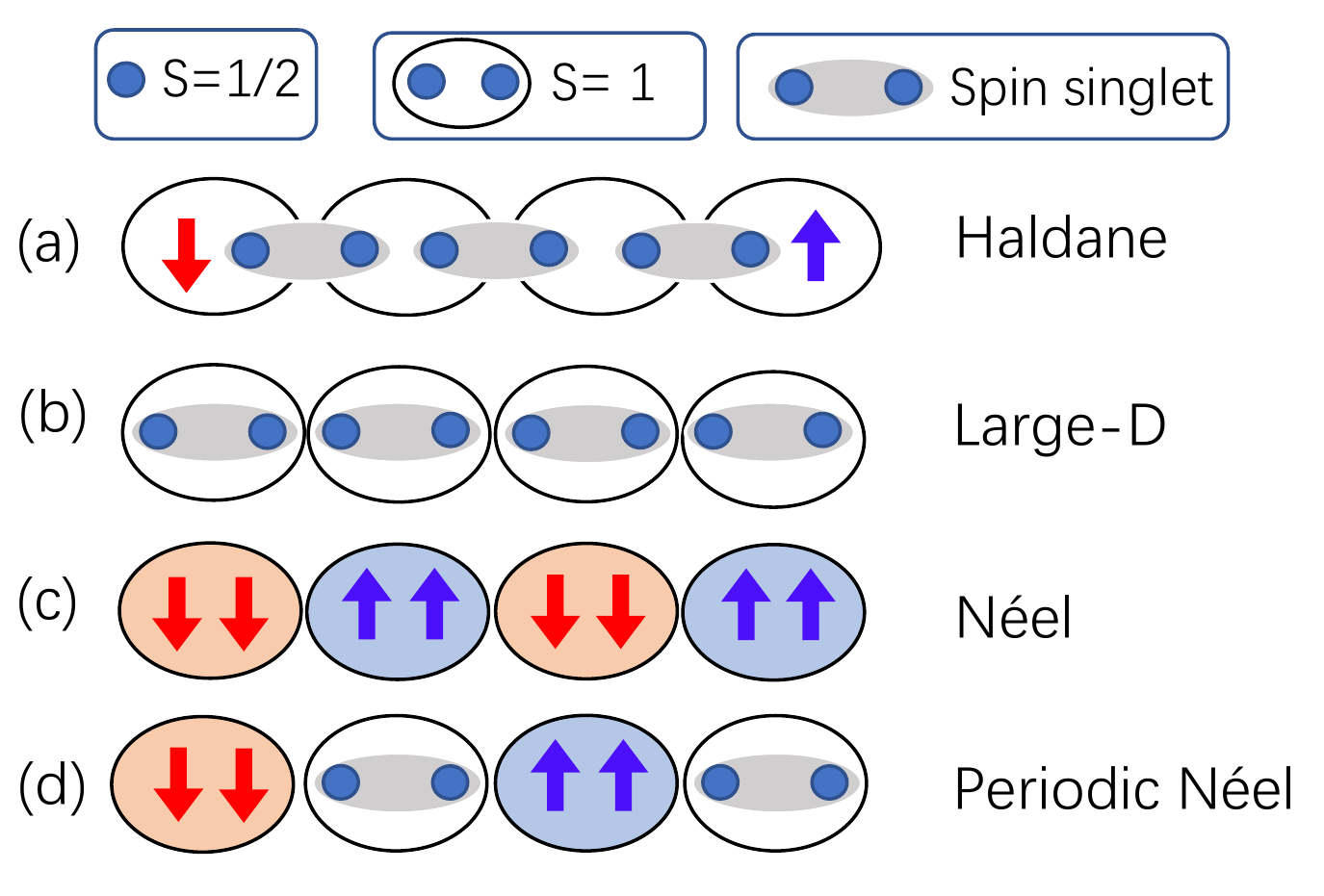}
\caption{
Schematic view of various phases considered in this paper for the 1D AF
$S=1$ chain. Each spin $S=1$ consists of two $S=1/2$ spins (blue dots);
a pair of spins built out of two adjacent spins and connected by a gray
ellipse forms a singlet,
\mbox{$(|{\uparrow\downarrow}\rangle-|{\downarrow\uparrow}\rangle)/\sqrt{2}$.}
Depending on actual parameters, the possible phases of a frustrated 1D
AF $S=1$ chain investigated here are:
(a) the Haldane phase (the AKLT state) which has two free edge
$S\!=\!1/2$ spins for an open boundary condition;
(b)~the singlet non-degenerate (large-$D$) state for a sufficiently large
single-site anisotropy $D_1>0$;
(c) the usual $S=1$ N\'{e}el phase,
$|{\uparrow\downarrow\uparrow\downarrow}\rangle$;
\mbox{(d)~the periodic} N\'{e}el phase which includes $|0\rangle$ states
as separating $|{\uparrow}\rangle$ and $|{\downarrow}\rangle$ states,
i.e., \mbox{$|{\uparrow}0{\downarrow}0\rangle$.}
\label{Phasesketch}}
\end{figure}

Frustration may be introduced in the AF spin-1 chain by other
interactions, and as a result novel types of order could emerge.
For instance, Affleck, Lieb, Kennedy, and Tasaki (AKLT) proposed
\cite{Aff87} an extended model of an $S=1$ chain with additional
biquadratic spin interaction, whose ground state is exactly solvable
and gives the phase depicted in Fig. \ref{Phasesketch}(a)]. The
fourfold ground state degeneracy arises due to two free edge $S\!=1/2$
spins for a chain with an open boundary condition and provides crucial
insight into symmetry-protected topological (SPT) order. Such a general
bilinear-biquadratic Hamiltonian leads to a dimerised spin-1 chain for
a range of parameters \cite{Aff87}. Note that the biquadratic spin-spin
interaction, $\propto({\bf S}_i\cdot{\bf S}_{i+1})^2$, arises naturally
within the $S=1$ Heisenberg model describing the strong coupling theory
of the iron pnictides \cite{Sta11,Yu12,Wan19}.

Yet, perhaps the simplest frustrated model for the AF spin-1 chain
includes both AF nearest neighbor and next-nearest neighbor
interactions---it has been shown \cite{Kol96} that the latter dilutes
the AKLT phase by singlet bond correlations \cite{Pix14}. Here we go
further and analyze the frustrated spin-1 chain with infinite range AF
correlations which decay algebraically in presence of the superposition
of uniform and alternating uniaxial single ion anisotropy. We show
below that such a model gives an interesting competition of spin
correlations with those encountered in the Haldane phase.

Experimentally the Haldane chain was most comprehensively studied via
inelastic neutron scattering in \mbox{$S=1$} chain materials, such
as SrNi$_2$V$_2$O$_8$~\cite{Bera13,Bera15} and
Ni(C$_2$H$_8$N$_2$)$_2$NO$_2$(ClO$_4$)~\cite{Delica91}. To reproduce
the experimental findings for real materials, additional terms have to
be added to the ideal Heisenberg Hamiltonian, such as exchange
anisotropy, bond alternation, or single-ion anisotropy. Anisotropy
effects can then significantly modify the ground state magnetic
properties. A~sufficiently strong easy plane single-ion anisotropy
$\propto D(S_i^z)^2$ induces a Gaussian quantum phase transition (QPT)
with central charge $c=1$ from the SPT state to a topologically trivial
large-$D$ phase
\cite{Tzeng02,Chen2003,tzeng2017,Jren2018,Hida2005,Batchelor,Roncaglia}
[cf. Fig. \ref{Phasesketch}(b)], such as the one encountered in
NiCl$_2$4SC(NH$_2$)$_2$(DTN)~\cite{Zapf06,Zvyagin07}.

Recently, artificial materials have been adopted to simulate quasi-1D
quantum materials in atomic, molecular, and optical system
\cite{Bloch2008,Bloch2012,Kim2010}. Different power-law decays of
long-range AF exchange interactions were considered in quasi-1D
quantum chains, such as the Coulomb-like interaction $\propto 1/r$
\cite{Saffman}, the dipole-dipole interaction $\propto 1/r^3$
\cite{Lahaye,Deng,Yan13}, and the van der Waals interaction
$\propto 1/r^6$~\cite{Saffman}. More recently, the power-law-decaying
long-range interaction $\propto 1/r^\alpha$ were also realized in which
the power $\alpha\ge 1$ could even be continuously adjusted in some
region using careful manipulation
\cite{Lahaye,Schneider,Browaeys,Britton2012,Islam,Gorshkov,Jurcevic14}.

We emphasize that the long-range interactions can not be considered
as perturbations and play essential role in the critical phenomena
\cite{W,Koffel,Sun01,Zhu,gong16,gong17,gong17L,gong16R,Frerot}.
Recently two of us considered a chain with anisotropic long-range
decaying interactions and investigated its QPTs \cite{Ren20}.
Interesting phase diagrams were established, with the long-range
interactions of the $z$ component resulting in a Wigner crystal phase,
and the transversal one resulting in a symmetry broken phase. The
present work addresses primarily the QPTs in the spin-1 chains with
long-range interactions and modulated single-ion anisotropy. Using a
combination of the density-matrix renormalization group (DMRG)
calculations and spin-wave analysis, various correlations for a spin-1
chain with long-range interactions were presented. The concept of
fidelity susceptibility has been successfully used to classify QPTs in
the spin-1 AF chain in the past
\cite{Tzeng01,Tzeng02,Ren09,Sch09,Alb10,Pol10,Dam11,Ren15,Su,Jren2018,Ram18}.
We employ here the fidelity susceptibility and establish a rich phase
diagram of spin-1 chain by changing the uniform and alternating
single-ion anisotropy.

The remainder of this paper is organized as follows. We introduce the
$S=1$ Heisenberg model with long-range interactions and alternating
single-ion anisotropy in Sec.~\ref{sec:Hamiltonian}. The details of
numerical methods and information metric are also introduced. In Sec.
\ref{sec:results}, effects of long-range interactions and modulated
single-ion anisotropy are investigated. Using fidelity susceptibility,
the phase diagrams for various cases are determined. The characteristics
of each phase are interpreted by diverse correlations and effective
Hamiltonians in various limiting cases. The discussion and summary are
presented in the last Sec.~\ref{sec:Discussion}.

\section{Hamiltonian and correlations}
\label{sec:Hamiltonian}

\subsection{Frustrated Heisenberg chain}
\label{sec:model}

The Hamiltonian of the spin-$1$ Heisenberg chain of length of $N$ sites
with long-range interactions and modulated single-ion anisotropy is
given by
\begin{equation}
\label{Hamiltonian}
H=\sum_{i<j}^N J_{ij}\textbf{S}_i\cdot\textbf{S}_j
+\sum_{i=1}^N\left[D_{1}+(-1)^iD_2\right]\left(S_i^z\right)^2,
\end{equation}
where $\textbf{S}_i$ is the spin-1 operator on site $i$. We consider
spin states using the following notation $|{\uparrow}\rangle$,
$|0\rangle$, and $|{\downarrow}\rangle$ for the single-site states with
$S_i^z=1$, 0, and $-1$, respectively. The interactions $J_{ij}$ between
two spins decay algebraically with distance \mbox{$r=|i-j|\ge 1$}, i.e.,
as
\begin{equation}
\label{Jij}
J_{i,i+r}=Jr^{-\alpha},
\end{equation}
and we take $\alpha\ge 1$. The parameters $D_1$ and $D_2$ stand for
uniform and alternating components of uniaxial single-ion anisotropy,
respectively. For convenience, we take $J=1$ for nearest neighbors
($|i-j|\equiv 1$) and use open boundary conditions in this study. Note
that for $D_1=D_2$ one finds single-ion anisotropy only at every second
site.

\subsection{Order parameter and spin correlations}
\label{sec:sop}

According to Ginzburg-Landau scenario, a well defined order parameter
is a vital ingredient for characterizing the nature of a phase.
In order to characterize the QPTs, the two-point correlations
\begin{eqnarray}
C^\alpha_{l,m}=\left\langle S^\alpha_l S^\alpha_m\right\rangle, \quad
( \alpha  = x, y, z),
\end{eqnarray}
and the nonlocal string order parameter (SOP),
\begin{eqnarray}
O^z_{l,m}=-\left\langle S^z_l \exp\left(
i\pi\sum_{k=l+1}^{m-1}S_k^z\right) S^z_m \right\rangle,
\label{sop}
\end{eqnarray}
are defined. The difficulty in defining suitable order parameters
in miscellaneous phases therein motivate us to adopt instead widely
accepted information measures.

\begin{figure*}[t]
\includegraphics[width=.99\textwidth]{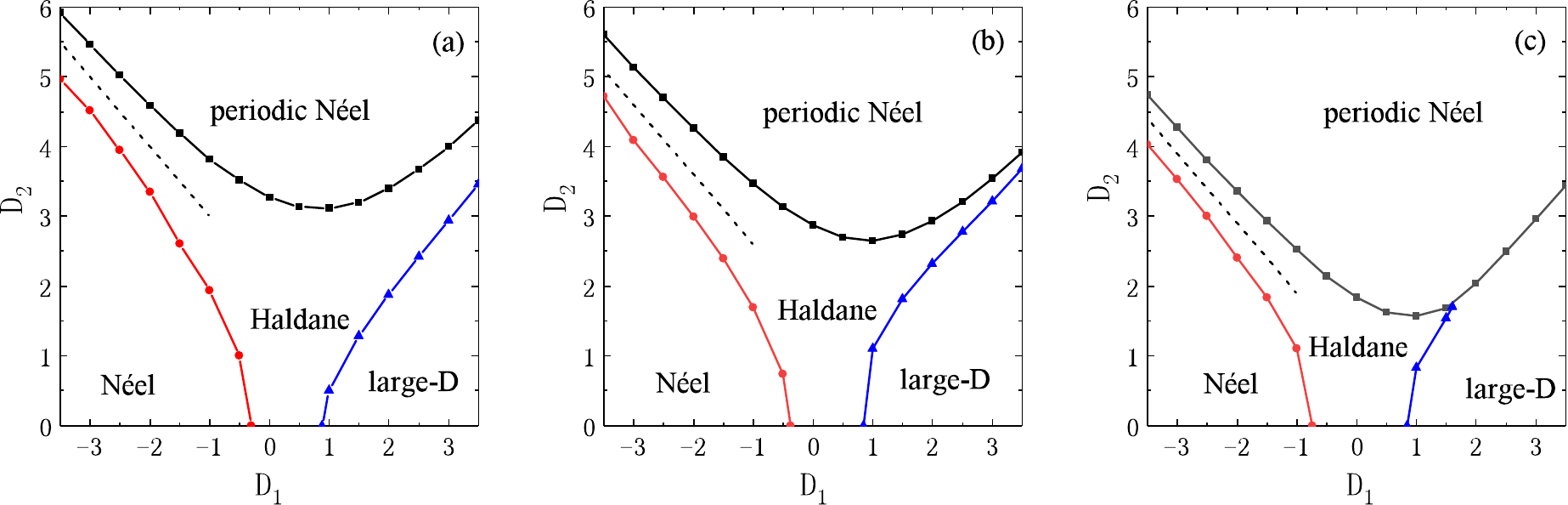}
\caption{Phase diagram of the spin-1 AF chain Eq. (\ref{Hamiltonian})
as in the $\{D_1,D_2\}$ plane with different $\alpha$. The location of
transition lines are obtained for the system size of $N=100$ sites,
where do not find any finite size effects.
The dashed lines in panels (a)--(c) separate approximately N\'eel from
periodic N\'eel phase; they are given by $D_2=-D_1+c_0$, with $c_0$
selected for a given value of~$\alpha$.
Parameters are:
\mbox{(a) $\alpha=10$, $c_0\simeq 2.0$;}
(b) $\alpha=3$, $c_0\simeq 1.6$;
(c) $\alpha=1$, $c_0\simeq 0.95$.
\label{phase_total}}
\end{figure*}

As a Riemannian metric in the parameter space, quantum fidelity
susceptibility is intimately related to quantum fluctuations and
dissipative responses of the system. Consider the Hamiltonian
$H({\bm \lambda})$ with a set of external parameters,
${\bm\lambda}=\{\lambda_1,\lambda_2,\cdots,\lambda_\kappa\}$, where
$\kappa$ is the dimension of the parameter space. The quantum geometric
tensor describes the geometric structure upon projecting the dynamics
onto the (non-degenerate) ground state $\psi_0({\bm\lambda})$, given by
\begin{equation}
\chi_{\mu,\nu} =\langle \partial_\mu \psi_0( {\bm \lambda})\left(
1-\vert\psi_0({\bm\lambda}\right)\rangle\langle\psi_0({\bm \lambda})
\vert)\vert\partial_\nu\psi_0({\bm \lambda})\rangle.
\label{chi}
\end{equation}
The imaginary part of the geometric tensor
$\textrm{Im}\{\chi_{\mu,\nu}\}$ is related to the Berry curvature, and
the real part $\textrm{Re}\{\chi_{\mu,\nu}\}$ is dubbed as quantum
fidelity susceptibility which measures the change rate of the distance
between the two closest states as a driving parameter $\lambda_\mu$ is
modulated \cite{You}. The quantum geometric tensor has been
experimentally extracted by using a superconducting qubit
\cite{Tan2019}, coupled qubits in diamond~\cite{MYu2020}, and
exciton-polaritons in a planar microcavity~\cite{Gianfrate19}.

As an information metric in the Hilbert space, quantum fidelity
susceptibility has a gravity dual with the spatial volume of the
Einstein-Rosen bridge in anti-de Sitter (AdS) space~\cite{Miyaji15}.
The sensitivity is greatly enhanced especially for the system at the
quantum criticality comparing with that away from the critical region
\cite{Braun18}. The divergence of $\chi_{\mu,\mu}$ (\ref{chi}) (in what
follows we use an abbreviation $\chi$ for this quantity) can directly
signal a QPT and locate the quantum critical points. Quantum fidelity
susceptibility has been proved to play a role of universal order
parameter in identifying the \mbox{QPTs \cite{Buonsante,You17,You15}.}

Both of these observations require us to calculate the ground state or
the reduced density matrix of the system. Based on matrix product
states, the finite-size DMRG technique was adopted
\cite{white,U01,U02}. In the numerics we keep up to $m=300$ eigenstates
during the procedure of basis truncation and the number of sweeps is
$n=50$. These conditions guarantee that the simulation converges
sufficiently fast and the truncation error is smaller than~$10^{-9}$.

\section{Numerical results}
\label{sec:results}

\subsection{Phase diagrams}
\label{sec:phd}

First, we performed calculations for varying uniaxial single-ion
anisotropy parametrized by $D_1$ and $D_2$, see Eq. (\ref{Hamiltonian}),
with fixed values of $\alpha=10$, 3, and 1, respectively. One finds then
generic phase diagrams, with the N\'eel and large-$D$ phase for negative
(positive) value of $D_1$ and the Haldane phase in between, see Fig.
\ref{phase_total}. Large (positive) value of $D_2$ induces periodic
N\'eel state. The characteristic spin states are visualized for
different phases in Fig.~\ref{Phasesketch}. One finds always four
phases; their range of stability changes when the exponent $\alpha$,
which describes the long-range decay, varies from $\alpha=\infty$
(the AF Heisenberg model with nearest neighbor AF interactions) to
$\alpha=1$. However, the intermediate Haldane phase shrinks as $\alpha$
decreases.

Remarkably for the AF Heisenberg model with nearest neighbor
interactions (not shown for a general case), numerical studies revealed
that the Haldane phase creeps in between the periodic N\'eel phase, the
N\'eel phase and the large-$D$ phase. This limit is reached with
increasing value of $\alpha$ from a finite value---then the Haldane
phase is squeezed, and a direct QPT from the large-$D$ phase to the
periodic N\'eel phase occurs for large $D_1$ through a $1^{st}$ order
QPT. The critical line of the periodic N\'eel to large-$D$ transition
will occur at $D_1=D_2$ for sufficiently large single-ion anisotropy
$D_1$.

\begin{figure}[t!]
\includegraphics[width=\columnwidth]{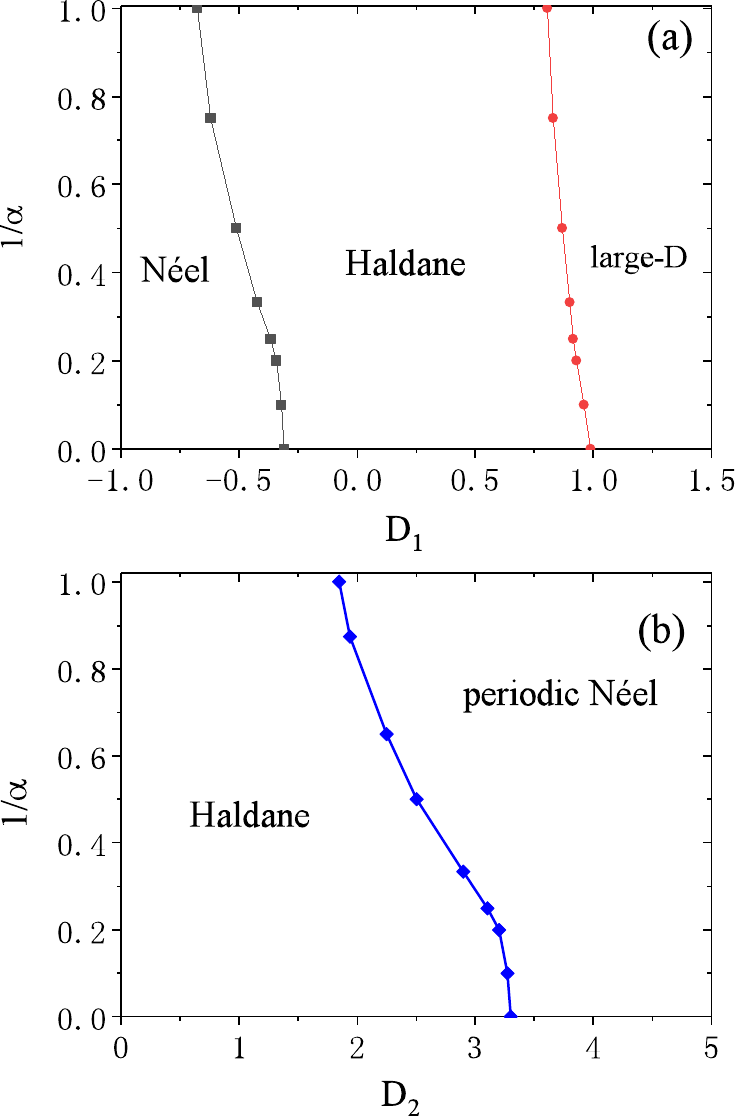}
\caption{Examples of QPTs found for the spin-1 chain in Fig.
\ref{phase_total} as functions of $\alpha$ for:
(a) uniform anisotropy $D_1$, with $D_2=0$, and
(b) modulated anisotropy $D_2$, with $D_1=0$.
\label{phase}}
\end{figure}

A tricritical point emerges at moderate $D_1$ for $\alpha=1$, see Fig.
\ref{phase_total}(c). However, there is still no direct transition
between the N\'eel phase and the periodic N\'eel phase, which emerges
from the Haldane state by lowering specific N\'eel-type spin
configurations compatible with the spatial modulation of single-site
anisotropy. A narrow Haldane phase would survive in the neighborhood of
the line $D_1+D_2=c_0$, i.e., at $D_2=-D_1+c_0$, where $c_0$ is a
constant and depends on the parameter $\alpha$.

To better understand such rich phase diagram of Fig. \ref{phase_total}
that results from the interplay of long-range AF couplings and
alternating single-ion anisotropies, we first consider the cases with
either $D_2=0$ or $D_1=0$, see Fig. \ref{phase}. For $\alpha=\infty$
(the AF Heisenberg model) ,
the system reduces to a spin-1 chain with nearest-neighbor interaction,
and it is in the Haldane phase at the isotropic point ($D_1=0$ and
$D_2=0$), which is composed of the superposition of states with hidden
nonlocal AF order. At $D_2=0$ one finds two QPTs by increasing the
value of $D_1$, from the N\'eel phase through Haldane phase to the
large-$D$ phase, see Fig. \ref{phase}(a). By going upwards vertically
at $D_1=0$, there is only one QPT for increasing $D_2$, from the
Haldane phase to the periodic N\'eel phase, see Fig. \ref{phase}(b).

\begin{figure}[t!]
\includegraphics[width=\columnwidth]{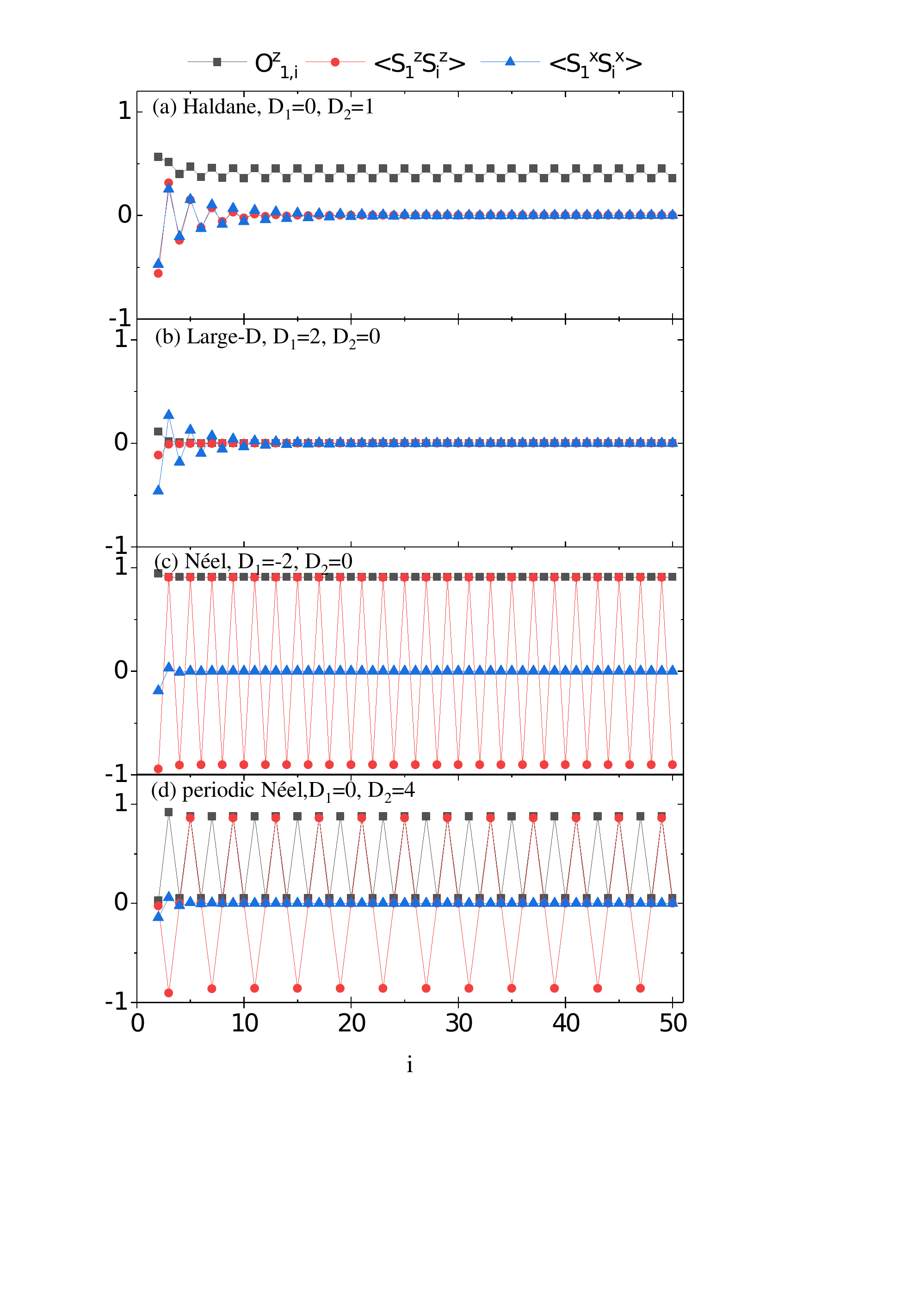}
\caption{\label{correlationr} Spin-spin correlation functions
$\langle S^\alpha_l S^\alpha_m\rangle$, and string order parameter
$O^z_{l,m}$ (\ref{sop}) as obtained with $N=100$ for four representative
sets of $\{D_1,D_2\}$ parameters (see legends) which correspond to
different phases of the spin-1 AF Heisenberg chain (with $\alpha=\infty$).
}
\end{figure}

Figure~\ref{correlationr} shows spin-spin correlation functions for the
different phases found in the phase diagram of Fig. \ref{phase_total}.
In Fig. \ref{correlationr}(a) one observes that both spin correlation
functions, $C^z_{1,i}$ and $C^x_{1,i}$, vanish fast over a separation
of just a few sites, while the nonlocal $O^z_{1,i}$ (\ref{sop})
converges quickly to 0.452 (0.358) for odd (even) site $i$, manifesting
the existence of the Haldane phase. The odd-even effect follows from
the open boundary condition. At $D_1=2$ and $D_2=0$ one finds the
large-$D$ phase, see Fig.~\ref{correlationr}(b).

Some specific AF order will be favored by varying the single-site
anisotropy. Changing the sign of $D_1$, one finds the N\'eel phase at
$D_1=-2$, see Fig.~\ref{correlationr}(c). The average magnetic moments
are: $\langle S_1^z\rangle\simeq 0.94$ for the edge site and
$\langle S_{50}^z\rangle\simeq 0.90$ in the middle of the chain. On
the other hand, for a quite large alternating single-site anisotropy,
i.e., $D_2\gg J$, the spins on the odd sites are restricted to
$\langle S_{2i-1}^z\rangle\approx\pm 1$ and those on the even sites are
confined to $\langle S_{2i}^z\rangle=0$. In Fig.
\ref{correlationr}(d) this state is shown for $D_1=0$ and $D_2=4$.
Indeed, two-spin correlations $C_{1,i}^z$ oscillate periodically
between the values being close to $-1$, 0, and 1, and
$C_{1,i}^x\approx 0$.

To elucidate the dominated configurations, the effective interactions
between $S_{2i-1}^z$ and $S_{2i+1}^z$ as the first-order perturbation
in $J$ plays the leading role, which for a chain of even length $N$ is
of the form,
\begin{eqnarray}
H_{\rm eff}^{(1)}=J_{13} \sum_{i=1}^{N/2-1} S_{2i-1}^z S_{2i+1}^z.
\end{eqnarray}
The crucial point here is the AF coupling $J_{13}$ between next nearest
neighbor spins at odd sites which triggers the periodic N\'{e}el phase,
being
$|...{\uparrow}0{\downarrow}0{\uparrow}0{\downarrow}\rangle...\rangle$
\mbox{[cf. Fig. \ref{Phasesketch}(d)];} the long-range interactions
frustrate this term. For an open system, one finds the net magnetization
being zero if $(N-2)$ sites is a multiple of 4, i.e., the considered
open chain may accommodate a certain number of the unit cells of length
four. Quantum fluctuation corrections are small but somewhat larger than
in the N\'eel phase, and one finds $|\langle S_{2i-1}^z\rangle|\simeq 0.88$.

\subsection{Spin correlations and fidelity susceptibility}
\label{sec:fide}

\begin{figure}[t!]
\includegraphics[width=0.99\columnwidth]{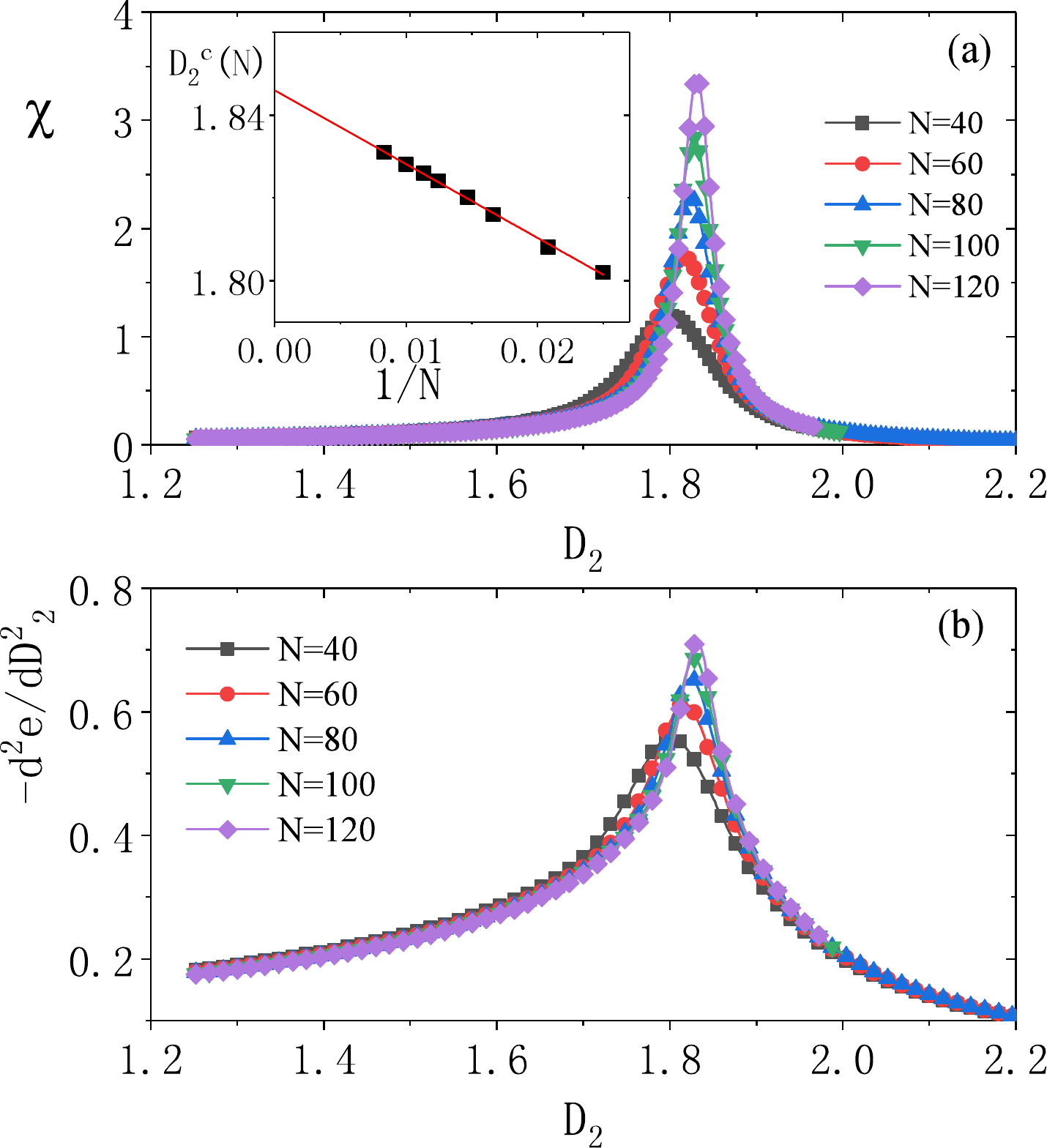}
\caption{(a) Fidelity susceptibility per site $\chi$ plotted as a
function of the alternating anisotropy parameter $D_2$ for different
system sizes (see legends). Inset shows the finite-size scaling of the
maximum of the fidelity susceptibility $D_2^{c1}$. The lines are fitted
and represent guides to the eye.
(b) The second derivative of the ground state energy density,
$(d^2e/dD_2^2)$, plotted for increasing parameter $D_2$ for different
system sizes. Parameters: $\alpha=1$ and $D_1=0$.
\label{Fidelity_s}}
\end{figure}

The phase transition between the Haldane phase and the periodic N\'{e}el
phase can be accurately determined by the analysis of the fidelity
susceptibility. The results for $D_1=0$ and $\alpha=1$ are shown in Fig.
\ref{Fidelity_s}(a). With increasing $D_2$, a peak of the ground-state
fidelity susceptibility is observed at $D_2^{*}\simeq 1.84$, which
signals approaching to the transition in the thermodynamic limit,
$N\to\infty$. Further evidence for identifying the QPT is provided by
the results of the second derivative of the ground-state energy density,
$(d^2e/dD_2^2)$, which is shown in Fig.~\ref{Fidelity_s}(b).
We thus confirm that the Haldane-to-periodic-N\'eel transition is a QPT
of $2^{nd}$ order.

According to the finite-size scaling theory~\cite{Fisher}, the position
of the maximal points of the fidelity susceptibility can be fitted by
the following formula:
\begin{equation}
|D_2^{*}(N)-D_2^c|\sim N^{-b},
\label{eq5}
\end{equation}
where $b$ is a constant given by the critical exponent $\nu$,
$b\equiv 1/\nu$, and $D_2^c$ is the quantum critical point in the
thermodynamic limit. Accordingly, the scaling of the extremal points of
the fidelity susceptibility gives rise to $D_2^c=1.845$, $b_1\simeq 1$,
as is shown in the inset of Fig.~\ref{Fidelity_s}(a). For a $2^{nd}$
order QPT, it is shown that the fidelity susceptibility at the peak
point for finite size $N$ behaves as
\begin{eqnarray}
 \chi(D_2^{*})\propto N^{\mu-1},
\end{eqnarray}
where $\mu$ is the critical adiabatic dimension.

\begin{figure}[t!]
\includegraphics[width=\columnwidth]{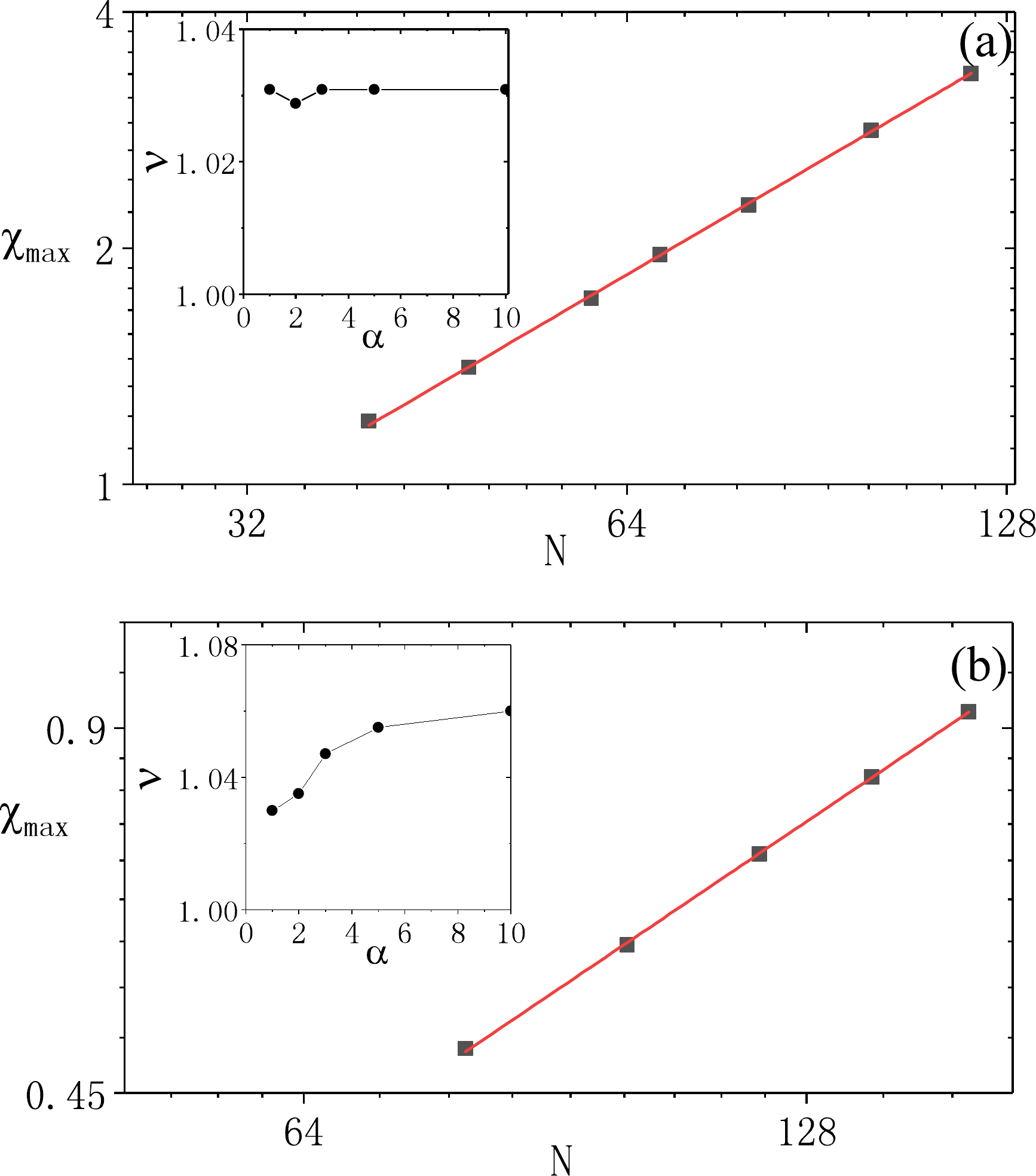}
\caption{Log-log plot of maximum fidelity susceptibility $\chi_{\rm max}$
as a function of the chain length $N$ with $\alpha=1$ for:
(a)~the~Haldane-to-periodic N\'{e}el QPT with $D_1=0$, and
(b)~the~Haldane-to-N\'{e}el QPT with $D_2=0$.
Insets show the critical exponent $\nu$ as a function of the parameter
$\alpha$, see Eq. (\ref{Jij}).
\label{Fidelity_Mu}}
\end{figure}

The critical exponent $\nu$ of the correlation length can be obtained
from $\mu$, $\nu=2/\mu$. As is disclosed in Fig. \ref{Fidelity_Mu}(a),
the linear dependence of log-log plot suggests that \mbox{$\mu=1.94$}
and $\nu=1.03$. This illustrates that the Haldane-to-periodic-N\'eel
QPT belongs to the Ising universality class \cite{Mikeska,You19}. We
recall that the phase diagram of the model Hamiltonian
(\ref{Hamiltonian}) for $D_1=0$ is shown in Fig. \ref{phase}(a), and
the whole critical line corresponds to $2^{nd}$ order QPT. For the AF
Heisenberg model (at $\alpha=\infty$), the Haldane-to-periodic-N\'eel
transition occurs at $D_2\simeq 3.30$ \cite{Hida2005,Ren15}. It is seen
that the critical point $D_2^{c1}$ drops when $\alpha$ decreases.

\begin{figure}[t!]
\includegraphics[width=0.99\columnwidth]{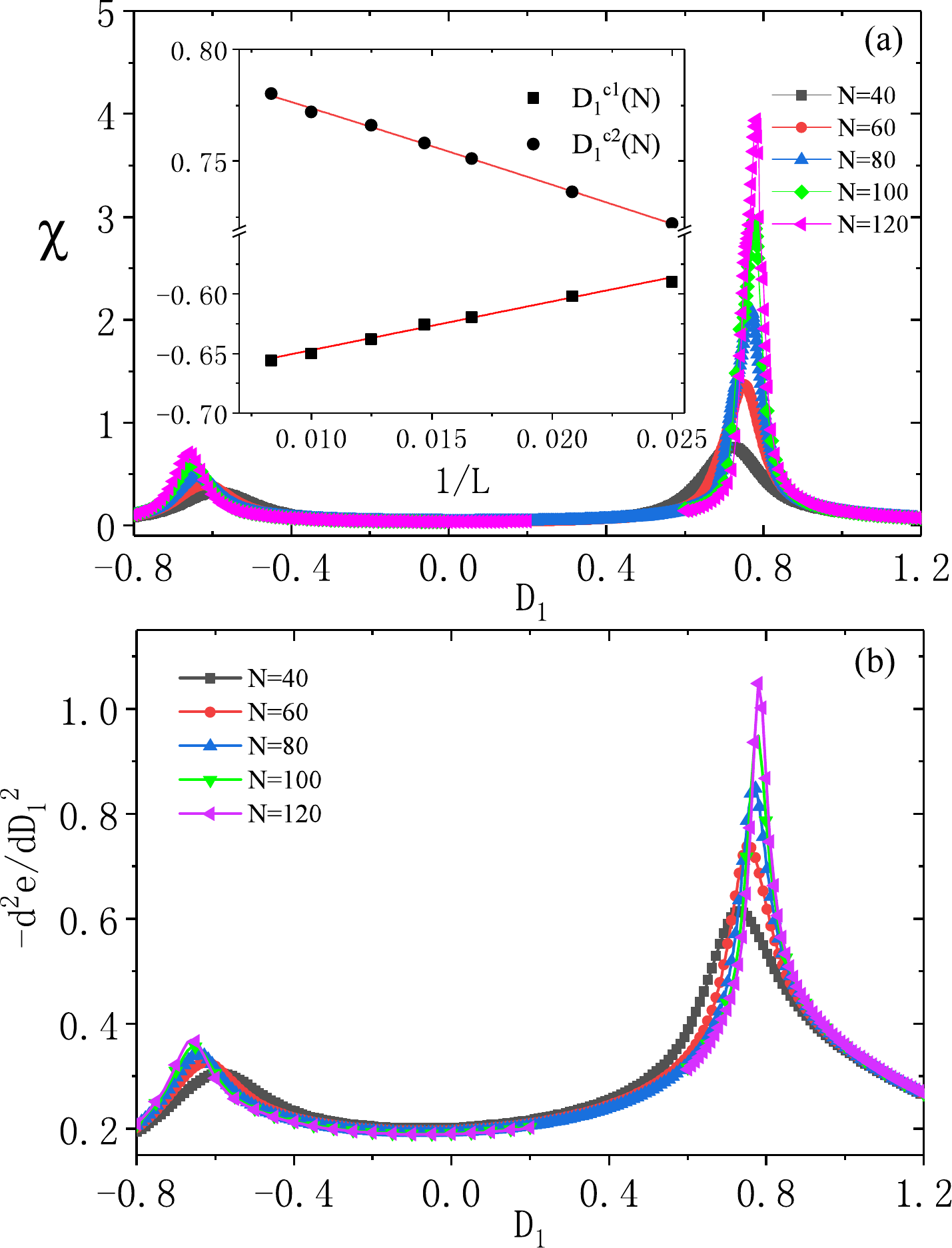}
\caption{
(a) Fidelity susceptibility $\chi$ per site plotted as a function of
the parameter $D_1$ for different system sizes, $N\in[40,120]$, with
$\alpha=1$. Inset shows the finite-size scaling of $D_{1}^{c1}$ and
$D_{1}^{c2}$ of the fidelity susceptibility. The lines are fitted
straight lines.
(b) The second derivative of the ground-state energy density plotted
as a function of the parameter $D_1$ for different system sizes,
$N\in[40,120]$, with $\alpha=1$.
\label{Fidelity}}
\end{figure}

Analogously, we consider the uniform case with $D_2=0$. Although the AF
long-range interactions are frustrated, the system would be reminiscent
of Haldane phase when $D_1=0$ and $\alpha\leq 1$ \cite{gong16}.
The system is in the Haldane phase for $\alpha=\infty$ and
$-0.31<D_1<0.99$ \cite{Tzeng01,Tzeng02,Chen2003,Roncaglia}.
For $D_1\ll -J$, all spins are restricted to be in the states
$\langle S_i^z\rangle=\pm 1$. When $\alpha=\infty$, the effective
coupling between the nearest~neighbor spins are obtained within the
first order perturbation theory in $J$ as~\cite{Hida2005},
\begin{eqnarray}
H_{\rm eff}^{(1)}=J_{12} \sum_i S_i^z S_{i+1}^z.
\end{eqnarray}
In this regard, the system is in the
$|{\uparrow}{\downarrow}{\uparrow}{\downarrow}\rangle$-type N\'eel
phase for a strong easy axis single-site anisotropy \mbox{$D_1<-0.31$.}
In the N\'eel phase, the two-spin correlations
\mbox{$C_{1,i}^z\simeq(-1)^{i-1}$} and $C_{1,i}^x$, decay exponentially
over the distance between two spins, as shown in Fig.
\ref{correlationr}(c). In the opposite limit, i.e., with $D_1\gg J$,
spins will be confined to be in the state
$\vert\langle S_i^z\rangle\vert=0$, annotating that the system enters
the large-$D$ phase for a sufficiently large easy plane anisotropy,
$D_1>0.99$. In the large-$D$ phase, the correlations $C_{1,i}^z$,
$C_{1,i}^x$, and $O^z_{1,i}$ vanish, see Fig.~\ref{correlationr}(b).

In order to determine the phase boundary with high accuracy, we also
calculated the fidelity susceptibility for $\alpha=1$ and $D_2=0$ of
the ground state for a system size $N$ up to $120$. The ground state
fidelity susceptibility per site $\chi_{D_1,D_1}/N$ is plotted as a
function of the parameter $D_1$ for different sizes $N$ in Fig.
\ref{Fidelity}(a). Two peaks in the fidelity susceptibility can be seen
and they both increase with increasing system size. This implies the
divergence of fidelity susceptibility in the thermodynamic limit, which
suggests the occurrence of two successive QPTs. We can conclude that
the left peak indicates the N\'eel-to-Haldane transition and the right
peak indicates the Haldane-to-large-$D$ transition. We then plot the
location of the maximum fidelity susceptibility as a function of $1/N$
and show the numerical fits in the inset of Fig. \ref{Fidelity}(a). We
obtain that $D_1^{c1}=-0.688$, $b_1\simeq 1.05$, and $D_1^{c2}=0.805$,
$b_2\simeq 1.00$ according to Eq. (\ref{eq5}).

\begin{figure}[t!]
\includegraphics[width=\columnwidth]{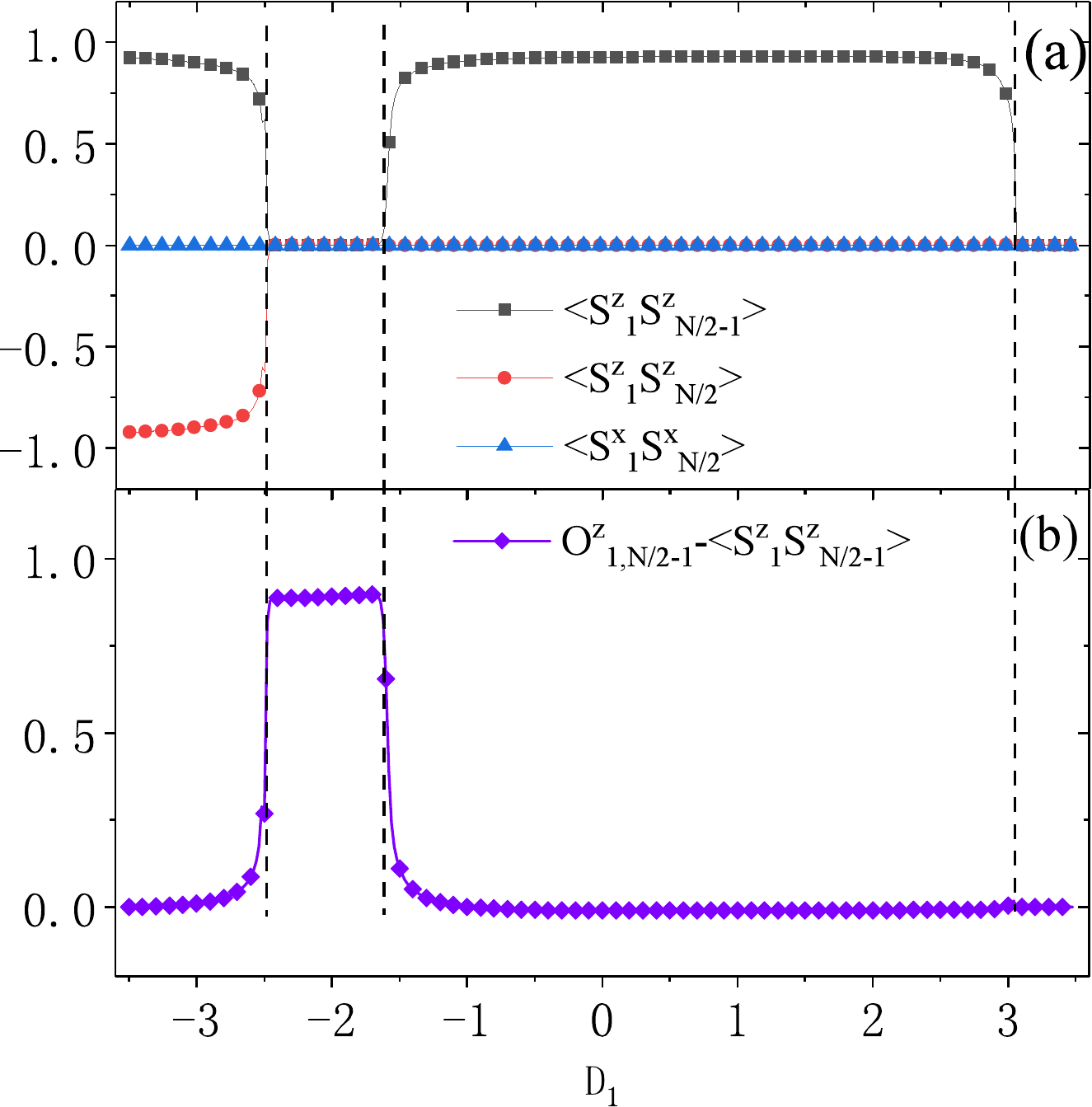}
\caption{\label{correlationD} Spin-spin correlation functions and
string order parameter are plotted as functions of $D_1$ with
$\alpha=1$, $D_2=3$. Parameter: $N=100$.}
\end{figure}

Further evidence which indicates QPTs is provided by the second
derivative of the ground state energy density $(d^2e/dD_1^2)$ in Fig.
\ref{Fidelity}(b). One observes that two peaks of $(d^2e/dD_1^2)$
become more pronounced for increasing system size $N$, which means that
both of the phase transitions are of $2^{nd}$ order. As is shown in
Fig. \ref{Fidelity_Mu}(b), the linear dependence of the log-log plot
gives rise to $\mu=1.93$ and $\nu=1.03$ for the Haldane-to-N\'eel QPT
with $\alpha=1$, $D_2=0$. The Haldane-to-N\'{e}el phase transition
belongs to the Ising universality class for all values of $\alpha$.
However, it was reported that the Haldane-to-large-$D$ transition may
be of higher order larger than two for the nearest neighbor model
($\alpha=\infty$)~\cite{Tzeng02}.
The long-range interaction would then reduce the order of the
transition. The values of critical points of Haldane-to-large-$D$ and
Haldane-to-N\'eel shift to a lower value when $\alpha$ decreases,
as is shown in Fig. \ref{phase}(b).

After considering the special cases, we then speculate that both the
easy-axis and easy-plane $D$-terms coexist in the Heisenberg chain.
The correlation functions and the SOP for $D_2=3$ are plotted as a
function of $D_1$ with $\alpha=1$ in Fig.~\ref{correlationD}(a).
One finds $C_{1,N/2-1}^z\simeq1$ and $C_{1,N/2}^z\simeq -1$ when
$D_1<-2.5$. After surpassing the critical point, the $z$-component
correlations vanish. When $D_1>-1.5$, the correlation $C_{1,N/2-1}^z$
rebounds but $C_{1,N/2}^z$ remains vanishing, which means that the
system enters the periodic N\'eel phase. Furthermore, $C_{1,N/2-1}^z$
vanishes suddenly at the periodic-N\'eel-to-large-$D$ transition point.

We also studied the SOP to characterize the phase in the range
$-2.5<D_1<-1.5$ in Fig.~\ref{correlationD}(b). It is noted that the SOP
$O^z_{1,i}\approx\vert C^z_{1,i}\vert$ in the N\'{e}el phase. In order
to determine the phase boundaries, we adopt
\mbox{$O^z_{1,N/2-1}-|C^z_{1,N/2-1}|$} to identify the Haldane phase,
in which the long-distance correlation $C^z_{1,i}$ disappears in the
Haldane phase~\cite{Ueda2008}.

\section{Conclusions }
\label{sec:Discussion}

In this paper, we investigated the quantum phase transitions in the
one-dimensional spin-$1$ chains with long-range antiferromagnetic
interactions decaying with a power law and modulated single-ion
anisotropy by using the density-matrix renormalization group technique.
Together with the short-range correlations and the nonlocal string
order parameter, the ground-state fidelity susceptibility was employed
to determine the phase diagram and critical phenomena.

The presence of long-range interactions increased the difficulty of
simulating the system numerically. Nevertheless, we provided compelling
evidence for the phase transitions and critical lines in the
thermodynamic limit. We identified four phases including the Haldane,
large-$D$, the
$|{\uparrow}{\downarrow}{\uparrow}{\downarrow}\rangle$ N\'eel phase,
and the $|{\uparrow}0{\downarrow}0\rangle$ periodic N\'eel phase.
The appearance of long-range interactions modifies the phase boundaries
and the order of phase transition comparing with their counterparts
with short-range interactions and leads to a direct $1^{st}$ order
transition between periodic N\'{e}el and large-$D$ phase. However,
a narrow Haldane phase survives between the periodic N\'eel and the
N\'eel phase as the long-range interactions increase.

In summary, employing the scaling functions of the fidelity
susceptibility gives a numerically economical way of obtaining
accurately the correlation-length critical exponents. We find that both
the Haldane-to-periodic-N\'eel and Haldane-to-N\'eel quantum phase
transitions are of $2^{nd}$ order and are classified for the model in
Eq. (\ref{Hamiltonian}) as belonging to the Ising universality class.
The local correlations are capable of characterizing those phases as
topological trivial ones, while only the nonlocal string order
parameter can identify the topological phase. More precisely, the
difference between the string order parameter and the two-point spin
correlation, $O^z_{1,N/2-1}-|C^z_{1,N/2-1}|$, can exclusively detect
the Haldane phase. We remark that the order of the Haldane-to-large-$D$
transition is higher than two for the nearest neighbor model
($\alpha=\infty$) and will drop to two as $\alpha$ decreases. Further
studies of the spin ordered phases at the presence of long-range
interactions is experimentally challenging and could also bring about
some novel types of phase transitions.

\begin{acknowledgments}

It is our pleasure to acknowledge insightful discussions with Marek
Rams and Gaoyong Sun. This work is supported by the National Natural
Science Foundation of China (NSFC) under Grants Nos.~11104021 and
11474211, as well as by the National Science Centre (NCN, Poland)
under Project No. 2016/23/B/ST3/00839.
W.-L. You acknowledges support by the startup fund (Grant No.
1008-YAH20006) of Nanjing University of Aeronautics and Astronautics.
A.~M.~Ole\'s is grateful for the Alexander von Humboldt Foundation
\mbox{Fellowship} \mbox{(Humboldt-Forschungspreis).}

\end{acknowledgments}

\end{document}